# Comparative Study Regarding Control of Wind Energy Conversion Systems Based on the Usage of Classical and Adaptive Neuro Fuzzy Controllers

**Iosif Szeidert**

Department of Automation and Industrial Informatics, Faculty of Automation and Computer Sciences, “Politehnica” University from Timisoara, Av. V. Parvan, No. 2, 300223, Timisoara, Romania, Phone: (0040) 256 403237, Fax: (0040) 256 403214, siosif@aut.utt.ro

*Abstract: The paper presents a comparative study regarding the control (using an adaptive neuro-fuzzy controller and a PD controller) based on the simulation of wind energy conversion systems functioning. There are considered several simulations based on asynchronous generator usage, by using the dedicated MATLAB-PSB (Power System Blockset) toolbox implementations.*

## 1 Introduction

The technical and scientific progresses, combined with the stimulation politics of ecological energy production and consumption conducted to a spectacular development of WECS (Wind Energy Conversion Systems). The economical-political worldwide trends indicate that the wind energy industry looks forward to a significant expansion.

The paper describes a WECS structure implemented in the MATLAB-Simulink simulation environment by using the specialized PSB toolbox, designed for modeling and simulation of energetic and electrical components. In Figure 1 there is presented the MATLAB - Simulink diagram of the considered windmill, which presents the following components: asynchronous machine, synchronous machine, wind turbine, frequency regulator and dump load. The practical problems in the grid integration of windmill represent special cases of design and analysis of energetic power systems. [2], [3], [5]

## 2 Wind Energy Conversion Systems – Control Structures Issues

The asynchronous machine model is based on the Park dq equations (detailed in appendix). The asynchronous machine operates in the generating regime. In this regime, the input into the machine is the mechanical energy and the output is the electrical energy. [1] The rotor speed must be adjusted in concordance to the changing wind speed in order to achieve the maximum aerodynamic efficiency. In this WECS, the wind turbine is modeled as simple controlled mechanical torque source that supplies the asynchronous generator. [7], [6] The frequency regulator input is represented by the voltage's frequency. There is used a three-phase Phase Locked Loop (PLL) system to measure the frequency of the 3-phase voltage of the network. Therefore, the measured frequency is compared to the reference frequency in order to obtain the frequency error. Afterwards, this error is integrated in order to obtain the phase. A classic PD type controller derives the error phase.

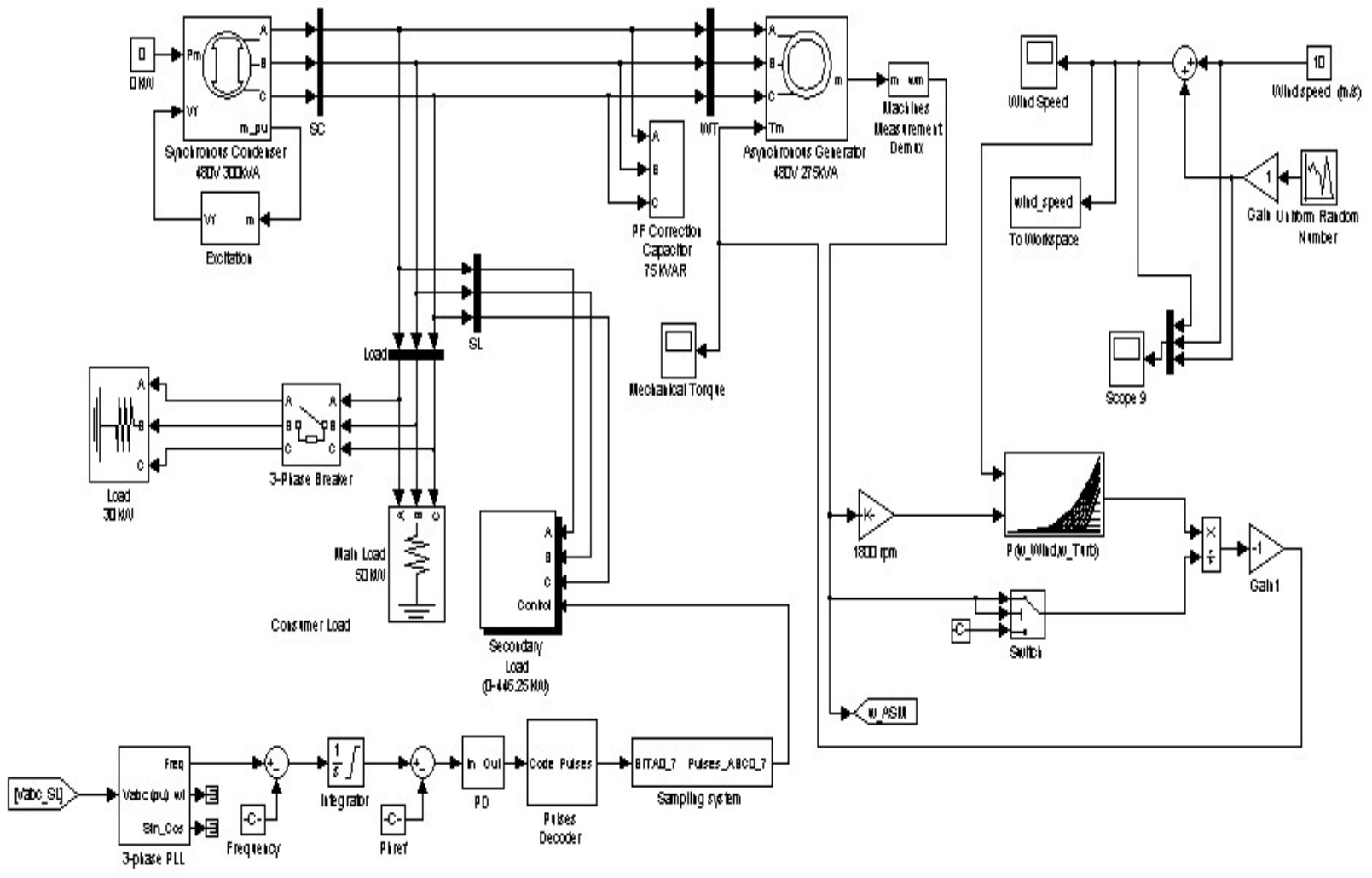

Figure 1

The MATLAB-Simulink WECS Diagram

The obtained signal (analogue signal) is afterwards converted to an 8-bit signal that commands the switching elements from the dump load. The regulator's output represents the desired power of the dump load. The dump load is used to dissipate the excess power produced by the windmill and simultaneously to maintain constant the frequency. The dump load consists of eight three-phase resistors connected in series with GTO (Gate Turn-Off) type based switches. The dump load uses an 8-bit binary command. [3]

In the considered windmill control structure, the synchronous machine presents only the role of a synchronous compensator. In Figure 2 there is detailed the structure of the considered frequency regulator. This structure has been used in the first study cases (SC1).

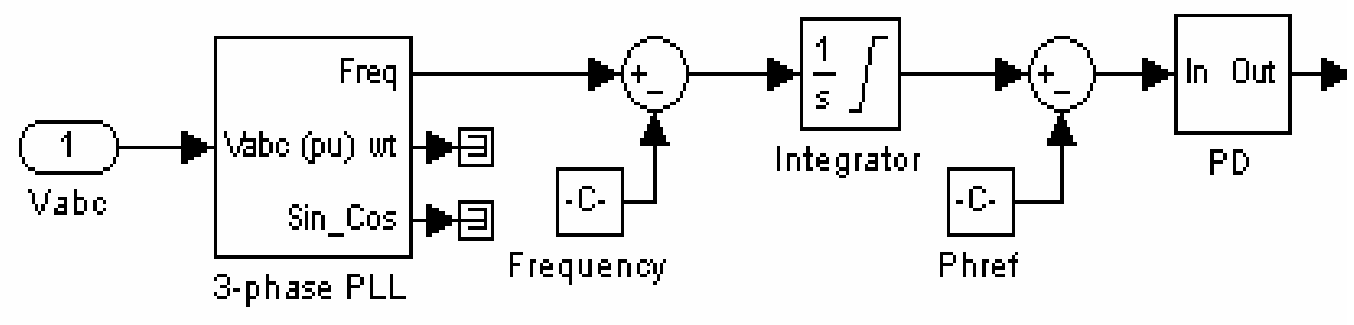

Figure 2
The frequency regulator

The second study case (SC2) was considered by using a modified control structure. The PLL block and conventional PD controller has been replaced with an adaptive neuro-fuzzy controller. The controller considers the prescribed frequency and phase reference values constant. The proposed modified structure is presented in Figure 3.

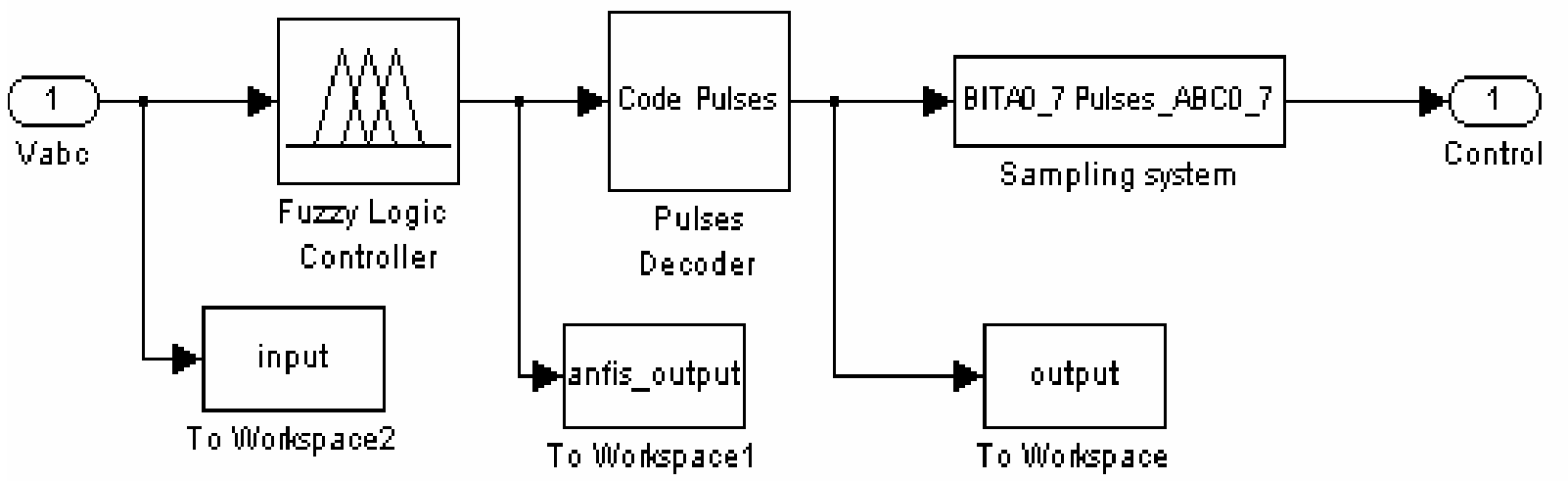

Figure 3
The frequency regulator based on an adaptive neuro-fuzzy controller

# 3 Wind Energy Conversion Systems Study Cases – Simulation Results

## 3.1 Study Case – SC1

### The Asynchronous Generator's Nominal Functioning Regime at Rated Power

The first study case is considered the normal functioning regime at rated power and rated wind speed. The wind speed is considered having the average value of 8 (m/s) during the simulation time period. The wind speed presents a random variation. The simulation interval was set to 3 seconds. There is connected only a main consumer of 50 kW. At $t_1$=1 [s] there is also connected a secondary consumer with nominal power of 40 kW. At $t_2$=2 [s] the secondary consumer is disconnected. This fact can be noticed in Figure 6, which represents the evolution of the considered load power. In Figure 4, is represented the evolution of the electrical energy frequency. It can be observed that in the condition of a continuous wind speed variation and of a diffrent load power the entire WECS control structure succeeds to control the electrical energy parameters (voltage and grid frequency). There can be noticed that the frequency presents only slightly variations. In Figure 5, is represented the asynchronous machine speed (rpm) in pu (per unit) units. There can be noticed that the rotation speed is slightly over the synchronous speed because the machine operates in generating mode. The wind turbine's power evolution (kW) is represented in pu units (Figure 7). It can be concluded that the considered WECS's control structure presents quite good control performances. [8], [9]

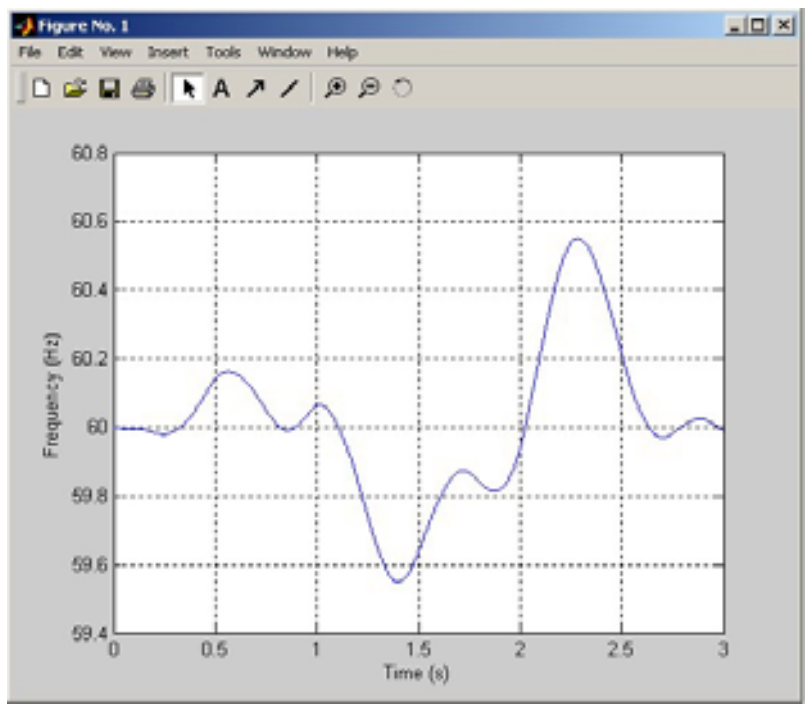

Figure 4
Frequency (Hz)

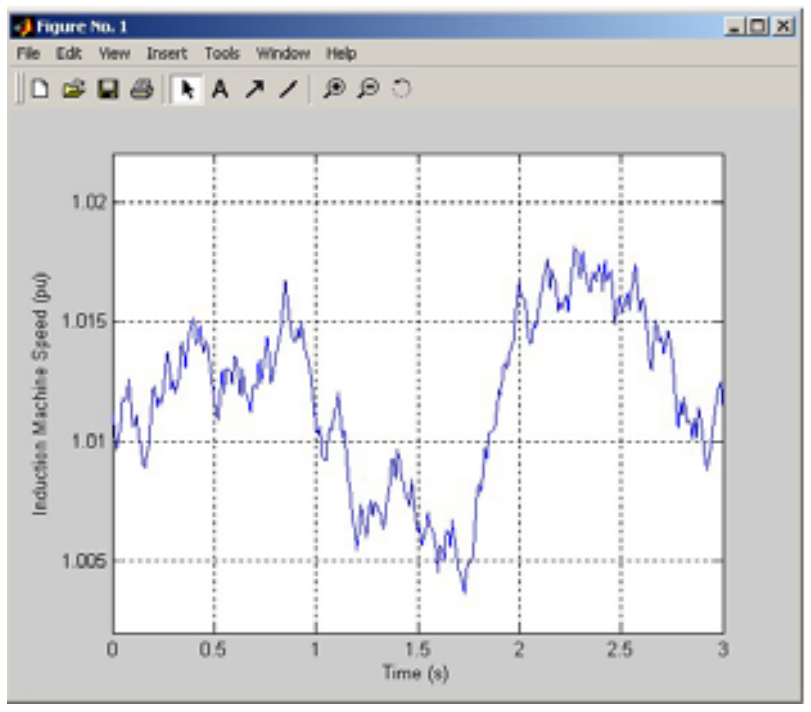

Figure 5
Asynchronous machine's speed (pu)

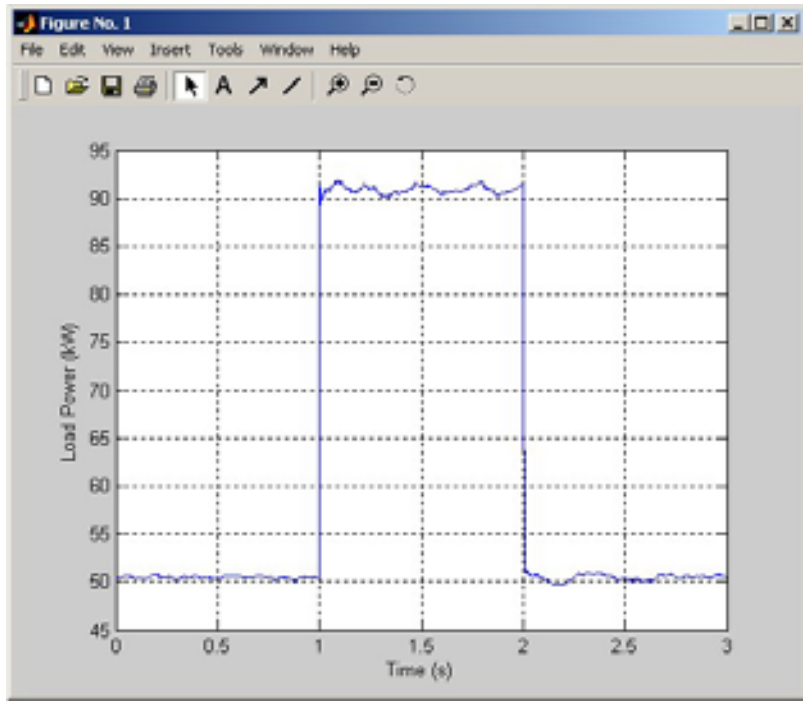

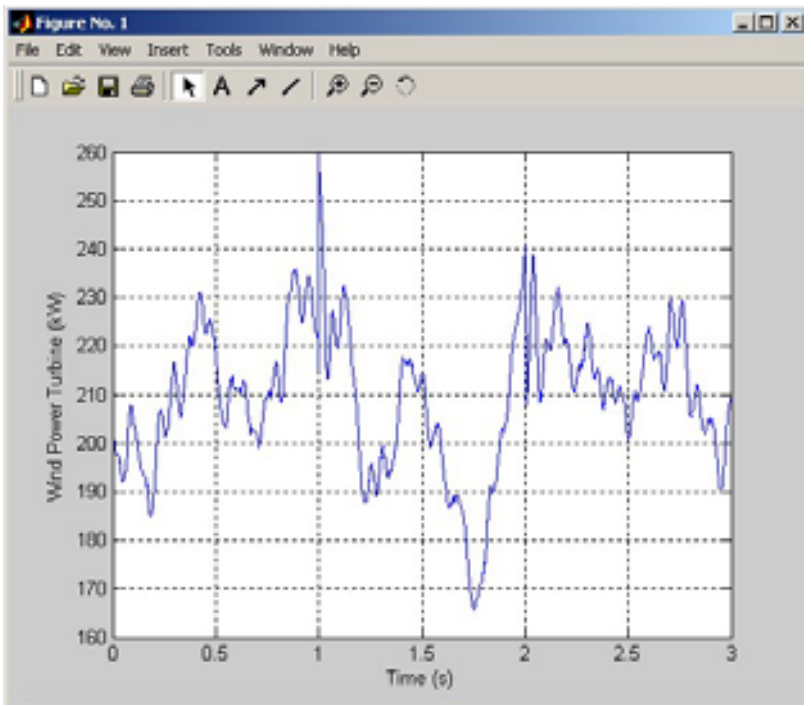

Figure 6
Load power (kW)

Figure 7
Wind turbine power (kW)

## 3.2 Study Case – SC2

**The Asynchronous Generator's Nominal Functioning Regime at Rated Power (Considering the Adaptive Neuro-Fuzzy Controller)**

In this study case there were considered the same simulation conditions as in the previous case (case SC1), but using the modified control structure based on an adaptive neuro-fuzzy controller. The obtained results are presented in Figures 8 and 9, the frequency and respectively, the asynchronous machine's speed. As it can be remarked the control structure presents similar performance level, which still satisfies the imposed goal of the windmill control. The modified control structure seems to represent a viable alternative control solution for this WECS structure.

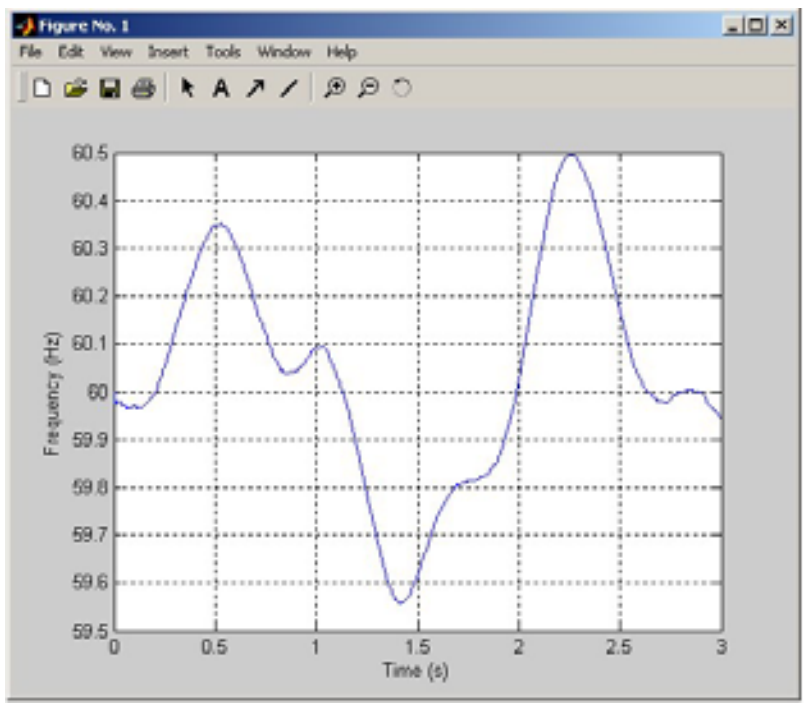

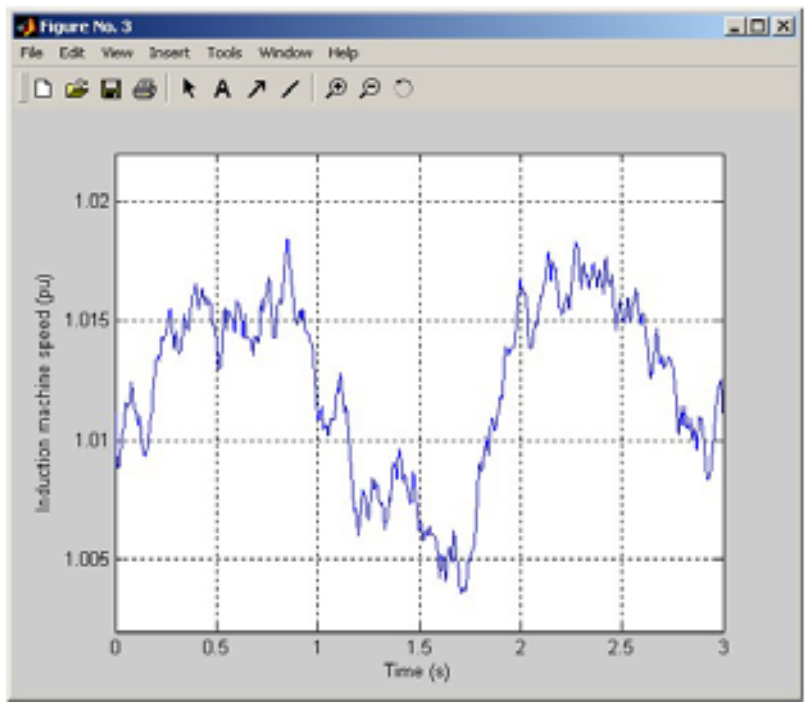

Figure 8
Frequency (Hz)

Figure 9
Asynchronous machine's speed (pu)

## Conclusions

There can be concluded that both controller structures present similar control performances. The usage of the adaptive neuro-fuzzy controller presents the advantage of an easier implementation on specialized computational structures, such as the case of control algorithms implemented on DSP processors or microcontroller based control systems.

The considered study cases present significance in the domain of the global renewal energies usage, especially in the case of WECS. However, in order to study thoroughly the WECS performances there are mandatory detailed analysis of the behavior of all main components of the WECS line: wind turbine, electrical generator, converter and grid. [4]

There can be concluded that in the considered study cases (nominal functioning regimes) the considered WECS structure presents an overall good performance regarding the imposed energy power parameters in the context of the integration into a distributed power grid (non-autonomous wind farms).

## Appendix

### Asynchronous machine model

The mathematical model of the asynchronous machine presents two subcomponents: the electrical system and the mechanical system. The equations (1)-(9) represent in fact the Park dq - asynchronous machine model:

Electrical system equations:

$$U_{qs} = R_S i_{qs} + \frac{d}{dt}(L_s i_{qs} + L_m i_{qr}) + \omega(L_s i_{ds} + L_m i_{dr}) \quad (1)$$

$$U_{ds} = R_S i_{ds} + \frac{d}{dt}(L_s i_{ds} + L_m i_{dr}) - \omega(L_s i_{qs} + L_m i_{qr}) \quad (2)$$

$$U_{qr} = R_r i_{qr} + \frac{d}{dt}(L_r i_{qr} + L_m i_{qs}) + (\omega - \omega_r)(L_r i_{dr} + L_m i_{ds}) \quad (3)$$

$$U_{dr} = R_r i_{dr} + \frac{d}{dt}(L_r i_{dr} + L_m i_{ds}) - (\omega - \omega_r)(L_r i_{qr} + L_m i_{qs}) \quad (4)$$

$$T_e = \frac{3}{2} p\left((L_s i_{ds} + L_m i_{dr}) i_{qs} - (L_s i_{qs} + L_m i_{qr}) i_{ds}\right) \quad (5)$$

Where:

$$L_s = L_{ls} + L_m \text{ and } L_r = L_{lr} + L_m \quad (6)\text{-}(7)$$

Mechanical system equations:

$$\frac{d}{dt}\omega_m = \frac{1}{2J}(T_e - T_m) \text{ and } \frac{d}{dt}\theta_m = \omega_m \quad (8)\text{-}(9)$$

Where: $R_s, L_{ls}$ - stator resistance and leakage inductance, $R_r, L_{lr}$ - rotor resistance and leakage inductance, $L_m$ - magnetizing inductance, $L_s, L_r$ - total stator and rotor inductances, $U_{qs}, i_{qs}$ - q axis stator voltage and current, $U_{qr}, i_{qr}$ - q axis rotor voltage and current, $U_{ds}, i_{ds}$ - d axis stator voltage and current, $U_{dr}, i_{dr}$ - d axis rotor voltage and current, $\omega_m$ - angular velocity of the rotor, $\theta_m$ - rotor angular position, $p$ - number of pole pairs, $\omega_r$ - electrical angular velocity ( $\omega_m \times p$ ), $\theta_r$ - electrical rotor angular position ( $\theta_m \times p$ ), $T_e$ - electromagnetic torque, $T_m$ - shaft mechanical torque, $J$ - combined rotor and load inertia constant.